\documentclass[aps,amssymb,amsmath,pra,reprint,noshowpacs, superscriptaddress]{revtex4-1}
\usepackage{times}
\usepackage{amssymb,amsmath,graphicx}
\usepackage[usenames]{color}
\usepackage{bbold}
\usepackage{siunitx}
\usepackage{braket}
\usepackage{comment}
\usepackage[T1]{fontenc}
\usepackage[utf8]{inputenc}
\usepackage[english]{babel}

\usepackage{hyperref}
\hypersetup{colorlinks=true, citecolor=blue, urlcolor=blue, linkcolor=blue}

\usepackage[normalem]{ulem}
\usepackage{lineno}

\newcommand{\Og}{\ensuremath{\Omega}}

\newcommand{\vect}[1]{\boldsymbol{#1}}
\newcommand{\tens}[1]{\tensor{#1}}

\newcommand{\imu}{\text{\rm i}}
\newcommand{\expu}{\text {\rm e}}
\newcommand{\diff}{\text{d}}
\newcommand*\colvec[3][]{
    \begin{pmatrix}\ifx\relax#1\relax\else#1\\\fi#2\\#3\end{pmatrix}
}

\let\baraccent=\= 

\newcommand{\re}[1]{\text{Re}\left[#1\right]}

\newcommand{\nx}{\vect{n}_x}

\newcommand{\nni}{\vect{n}_i}
\newcommand{\nphi}{\vect{n}_\phi}

\newcommand{\ntheta}{\vect{n}_\theta}
\usepackage{textcomp}

\newcommand{\commentOut}[1]{}
\usepackage{scalefnt}   

\newcommand{\affil}{Photonics Laboratory, ETH Zürich, CH-8093 Zürich, Switzerland}
\newcommand{\affilft}{Department of Physics, Humboldt-Universit{\"a}t zu Berlin, 10099 Berlin, Germany}
\newcommand{\affilan}{Institute of Photonics, University of Eastern Finland, P.O. Box 111, FI-80101 Joensuu, Finland}
\newcommand{\affilQC}{Quantum Center, ETH Zürich, CH-8093 Zürich, Switzerland}

\begin{document}
\scalefont{1.05}
\title{Optimal orientation detection of an anisotropic dipolar scatterer}

\author{Felix \surname{Tebbenjohanns}}\affiliation{\affil}
\affiliation{\affilft}
\author{Andrei \surname{Militaru}}\affiliation{\affil}
\author{Andreas \surname{Norrman}}\affiliation{\affil}
\affiliation{\affilan}
\author{Fons \surname{van der Laan}}\affiliation{\affil}
\author{Lukas \surname{Novotny}}\affiliation{\affil}
\affiliation{\affilQC}
\author{Martin \surname{Frimmer}}\affiliation{\affil}
\homepage{http://www.photonics.ethz.ch}


\date\today

\begin{abstract}
    The angular orientation of an anisotropic scatterer with cylindrical symmetry in a linearly polarized light field represents an optomechanical librator. 
    Here, we propose and theoretically analyze an optimal measurement scheme for the two angular degrees of freedom of such a librator. 
    The imprecision-backaction product of this scheme reaches the Heisenberg uncertainty limit. Furthermore, we propose and analyze a realistic measurement scheme and show that, in the absence of spinning motion around the symmetry axis, measurement-based ground-state cooling of the rotational degrees of freedom of an anisotropic point scatterer levitated in an optical trap is feasible.
\end{abstract}

\maketitle

\section{Introduction}
At the heart of optomechanics lies the task to measure and control mechanical motion using light~\cite{aspelmeyer2014cavity}. To a large extent, the optomechanics community has focused on translational degrees of freedom, such as the position of a mirror~\cite{Abbott2016}, of a nanomechanical membrane~\cite{Rossi2018}, or of a particle in an optical trap~\cite{Millen2020}.
Recently, rotational degrees of freedom of anisotropic particles levitated in optical and radiofrequency traps have attracted significant attention~\cite{Reimann2018,Ahn2018,Ahn2020,Bang2020,Delord2020}.
A dumbbell, composed of two identical spherical particles in touching contact, is an example of an anisotropic scatterer.
In a linearly polarized laser field, dumbbells align with their long axis along the polarization direction due to the exerted optical torque. 
For small angular deviations around the equilibrium position, this angular motion represents a harmonic oscillator degree of freedom, termed libration. Gaining quantum control of these libration modes is an exciting prospect~\cite{Stickler2021,Schaefer2021,Rudolph2021,Zhong2017,Seberson2019}, since it may allow the investigation of quantum coherent evolution of the rotational degrees of freedom of macroscopic objects~\cite{Stickler2016,Stickler2018,Stickler2018rotationquantum}. Accordingly, a quantum toolbox of rotational motion may provide an avenue that complements current efforts to investigate macroscopic superposition states using the center-of-mass motion of levitated particles~\cite{Romero-Isart2017}. Another prospect of optically levitated rotors is to harness them as torque sensors to investigate elusive physics of rotating bodies, such as the rotational Casimir effect and vacuum friction~\cite{Manjavacas2010,Zhao2012,Xu2017,Manjavacas2017}. Developing levitated librators into a quantum resource can take inspiration from the measurement and control schemes developed for center-of-mass modes~\cite{Magrini2021,Tebbenjohanns2021}. One of the crucial ingredients for quantum control of these modes has been the optimization of their optical detection~\cite{Tebbenjohanns2019Efficiency}. While first steps to measure and feedback-control the libration modes of levitated nanoparticles have been taken~\cite{Bang2020,vanderLaan2020shotnoise}, an understanding of the optimal detection process for the orientation of an optically levitated dumbbell is still missing.

In this paper, we analyze the detection of the librational motion of a lossless dipolar point scatterer with cylindrical symmetry (such as a dielectric nanodumbbell) in a linearly polarized laser field.
First, we propose and theoretically analyze an ideal detection system to measure both angular degrees of freedom. We show that the imprecision-backaction product of our measurement scheme reaches the limit set by the Heisenberg uncertainty relation and is thus optimal. Second, we propose and analyze a realistic detection scheme that has the potential to allow for measurement-based ground-state cooling of librational motion.

\section{System under investigation}\label{sec:system}
Throughout this work, we consider an absorption-free anisotropic dipolar point scatterer with cylindrical symmetry described by its polarizability tensor $\tens\alpha$. For a scatterer with vanishing material loss in the optical frequency range of interest, the imaginary part of the polarizability (which is due to radiation loss) is negligible for our purposes, such that $\tens\alpha$ can be taken as purely real. 
In a frame of reference with the space-fixed Cartesian $x$ axis aligned with the scatterer's body-fixed symmetry axis, the polarizability takes the form $\tens\alpha=\alpha_0\text{diag}[1,1-\Delta,1-\Delta]$~\cite{Zhong2017}. 
Here, $\alpha_0$ is the polarizablity along the scatterer's symmetry axis, while $\Delta<1$ characterizes the anisotropy of the scatterer.
An example of such a scatterer of particular practical relevance is the workhorse of rotational optomechanics, a dumbbell composed of two touching subwavelength spheres, as illustrated in Fig.~\ref{fig:idealScheme}(a). The scatterer is irradiated by a light field (angular frequency $\omega_0$), which is linearly polarized along $x$ at the location of the scatterer and has  amplitude $E_0$. We note that only the driving field at the scatterer position is of relevance and the mode shape of the driving field plays no role. The scatterer experiences a torque aligning its long axis with the polarization direction of the field. 
In this aligned situation, the dipole moment induced in the scatterer by the driving field points purely along the $x$ direction and is given by $p_x=\alpha_0 E_0$.
We are interested in measuring the angular deviation of the scatterer's orientation from this equilibrium position. 

For small angular deviations, it is sufficient to consider rotations around the space-fixed Cartesian $y$ and $z$ axes. Accordingly, the orientation of the scatterer is fully described by two angles $\delta$ and $\epsilon$, where $\delta$ denotes a rotation of the scatterer around the $z$ axis, and $\epsilon$ around the $-y$ axis, as illustrated in Fig.~\ref{fig:idealScheme}(b). 
Note that a third rotation around the $x$ axis does not appear due to the symmetry of the scatterer.
For small angles $\delta$ and $\epsilon$, the order in which these rotations are effected is irrelevant.
The induced dipole moment (in the space-fixed Cartesian frame) is then $\vect p = R_zR_{-y} \tens\alpha R_yR_{-z} E_0 \vect n_x$ with the rotation matrices $R_z$ (around the $z$ axis by $\delta$) and $R_y$ (around the $y$ axis by $\epsilon$), and the Cartesian unit vector $\vect n_x$. 
To linear order in the angles $\delta$ and $\epsilon$, we find $\vect p = [1, \Delta\delta, \Delta\epsilon] \alpha_0E_0$.
With $p_x$ the component of the dipole moment along the polarization direction, we can thus express the remaining components of the dipole moment as $p_y=\Delta\delta p_x$ and $p_z=\Delta\epsilon p_x$.

\section{Ideal measurement system}\label{sec:idealMeasurement}
The central question answered in this paper is: how can we optimally infer the orientation of the anisotropic scatterer (i.e., the angles $\delta$ and $\epsilon$) in a linearly polarized electromagnetic field?
In other words, we seek to determine the orientation of a radiating dipole with dipole moment $\vect{p}$ by analyzing its radiation field.

\begin{figure}[t]
\includegraphics[width=\columnwidth]{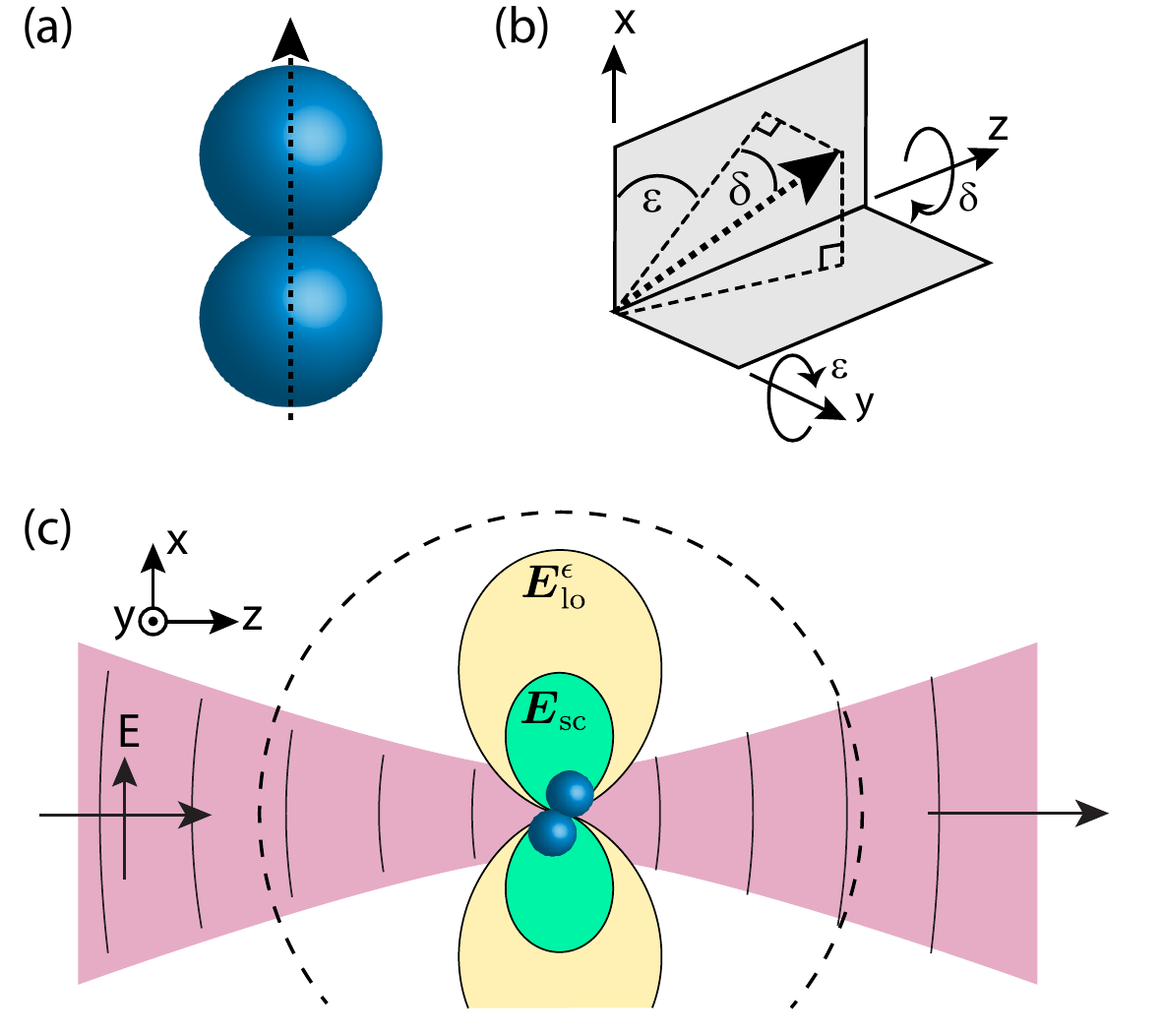}
\caption{(a)~Asymmetric scatterer with cylindrical symmetry depicted as a dumbbell composed of two touching spheres. The orientation of the dumbbell is described by the orientation of its axis of symmetry, illustrated by the dashed arrow. (b)~Illustration of the coordinate system. The space-fixed laboratory frame is described by the Cartesian coordinate axes $x$, $y$, and $z$. 
The orientation of the dumbbell (shown as the dashed arrow) can be described by the orientation of its symmetry axis relative to the space-fixed Cartesian axes. 
Small deviations of this symmetry axis from the Cartesian $x$ axis can be described by the rotation angle $\delta$ around the $z$ axis, and the rotation angle $\epsilon$ around the $-y$ axis.
(c)~Illustration of optimal detection scheme for the angle $\epsilon$. The dumbbell is driven by an $x$-polarized field, exemplarily illustrated in red as a beam of light traveling along the $z$ axis. The tilt by $\epsilon$ gives rise to an induced dipole moment along $z$. The field $\vect{E}_\text{sc}$ generated by that $z$-oriented dipole (illustrated in green) populates the mode $\vect{u}^{(z)}$. For optimal detection, that scattered field is mixed with a strong local oscillator field $\vect{E}_\text{lo}^\epsilon$ in the same dipolar radiation mode (illustrated in yellow). The field intensity is measured on a $4\pi$ detector, illustrated as the dashed black line.
}
\label{fig:idealScheme}
\end{figure}

In a rotating frame at the optical laser frequency $\omega_0$, we can write the complex electric far-field generated by the dipole $\vect{p}=[p_x,p_y,p_z]$ in spherical space-fixed coordinates as~\cite{Novotny2012}
\begin{equation}\label{eq:scattered_field_diople}
    \vect E_\text{sc}(r,\theta,\phi)= \frac{\omega_0^2}{\epsilon_0 c^2} \frac{\expu^{\imu k r}}{4\pi r} \tensor{G} \vect{p}, 
\end{equation}
where $\epsilon_0$ is the vacuum permittivity and $c$ the speed of light in vacuum. We denote the column vectors of the Green's tensor $\tensor{G}$ 
as $\tensor{G}\cdot \vect n_i = \sqrt{8\pi/3}\, \vect u ^{(i)}$, where $\vect n_i$ are the (space-fixed) Cartesian unit vectors ($i\in\{x,y,z\}$).
The polarization $\vect u ^{(i)}$ of the field emitted by a dipole oriented along Cartesian axis $i$ is given by~\cite{Novotny2012} 
\begin{equation}
    \vect u^{(i)} = \sqrt{\frac{3}{8\pi}} \left[ \left(\nni\cdot \ntheta\right)\ntheta + \left(\nni\cdot\nphi\right)\nphi \right],
\end{equation}
where 
$\ntheta$
is the space-fixed unit vector along the polar direction (relative to the $z$ axis) and
$\nphi$
the one along the azimuthal direction.
These dipolar modes are orthogonal in the sense that 
\begin{equation}\label{eq:orthogonality}
    \int\diff\Omega~  \vect u^{(i)} \cdot \vect u^{(j)} = \delta_{ij},
\end{equation}
where the integral is over the full solid angle and $\delta_{ij}$ denotes the Kronecker delta.
Integrating the intensity $(\epsilon_0 c /2)|\vect E_\text{sc}|^2$ over the surface of a sphere yields the total scattered power ${P_0=\omega_0^4\alpha_0^2 E_0^2/(12\pi\epsilon_0 c^3)}$, to linear order in $\epsilon$ and  $\delta$. In the following discussion, it is sufficient to analyze the fields on the surface of a sphere centered at the origin with some large radius $R$. 
To ease our notation, we re-normalize the electric field on this reference sphere and define 
\begin{equation}\label{eq:field_Dipole}
\begin{split}
    &\vect{\mathcal E}_\text{sc}(\theta, \phi) := \sqrt{\frac{\epsilon_0 c}{2}} R \expu^{-\imu k R} \vect E_\text{sc}(R, \theta, \phi)\\
    &= \sqrt{P_0}\left(\vect u^{(x)} + \Delta\delta \vect u^{(y)} + \Delta\epsilon \vect u^{(z)} \right).
\end{split}
\end{equation}
Integration of this field's modulus squared over the full solid angle yields the total scattered power ${P_0 = \int\diff\Omega~|\vect{\mathcal E}_\text{sc}|^2}$. 
Throughout this manuscript, electric fields (in V/m) are denoted by $\vect E$ while the re-normalized fields (in $\sqrt{\rm W}$) are denoted by $\vect{\mathcal E}$.

According to Eq.~\eqref{eq:field_Dipole}, the information about the angles $\delta$ and $\epsilon$ is encoded in the amplitudes of the $\vect{u}^{(y)}$ and $\vect{u}^{(z)}$ dipolar modes, respectively. 
In order to extract these amplitudes, and infer the scatterer's orientation, we use the homodyne detection scheme illustrated in Fig.~\ref{fig:idealScheme}(c). In this scheme, we combine the signal field $\vect{ \mathcal E}_\text{sc}$ with a local oscillator (LO) field, with an amplitude much larger than that of the signal. 
In App.~\ref{app:ideal_ref_field}, we show explicitly that an LO field in the same mode as the signal field maximizes the signal-to-noise ratio on the detector. 
Thus, we choose
\begin{equation}\label{eq:ideal_ref_field}
    \vect{ \mathcal E}_\text{lo}^\delta = \sqrt{P_\text{lo}} ~ \vect u^{(y)}
\end{equation}
as our LO field for measuring $\delta$ and $\vect{ \mathcal E}_\text{lo}^\epsilon = \sqrt{P_\text{lo}} ~ \vect u^{(z)}$ for measuring $\epsilon$.
To keep our discussion of the ideal measurement scheme accessible, we consider a measurement of either $\delta$ or $\epsilon$ at one time. Nevertheless, we stress that one can, in principle, measure both angles simultaneously, since their associated radiation modes are orthogonal according to Eq.~\eqref{eq:orthogonality}.

The power measured by our detector dedicated to the angle $\delta$ then reads
\begin{equation}\label{eq:Felix_P_det_delta}
    P_\text{det}^\delta = \int \diff\Omega ~ |\vect{ \mathcal E}_\text{lo}^\delta + \vect{ \mathcal E}_\text{sc}|^2 =  P_\text{lo} + 2\sqrt{P_\text{lo}P_0} \Delta \delta,
\end{equation}
where we assumed $P_0 \ll P_\text{lo}$.
The signal $P_\text{det}^\delta$ indeed provides a linear measurement of the orientation angle $\delta$, as desired. An analogous calculation for $P_\text{det}^\epsilon$ yields an expression identical to Eq.~\eqref{eq:Felix_P_det_delta}, with $\delta$ replaced by $\epsilon$. Assuming that the noise on our detector is dominated by photon shot noise with power spectral density $S_{PP} = \hbar\omega_0P_\text{lo}/(2\pi)$ (with $\hbar$ the reduced Planck constant and $\omega_0$ the optical frequency)~\cite{bowen2015quantum}, we find the measurement imprecision 
\begin{equation}\label{eq:S_epsepsS_deltadelta}
    S_{\delta\delta} = S_{\epsilon\epsilon} = \frac{1}{2\pi} \frac{1}{\Delta^2} \frac{\hbar\omega_0}{4P_0}.
\end{equation}
Here $S_{\delta\delta}$ and $S_{\epsilon\epsilon}$ denote the power spectral densities (PSD) of the measurements of $\delta$ and $\epsilon$, respectively~\footnote{\label{fn:psd_normalization}We normalize the PSD $S_{xx}(\omega)$ of a signal $x(t)$ such that $\braket{x(t)^2}=\int_{-\infty}^\infty\diff\omega~S_{xx}(\omega)$.}.
In App.~\ref{app:ideal_ref_field}, we provide a generalized treatment for the measurement imprecision of any linear measurement limited by photon shot noise.
There, we provide a quantum mechanical derivation of $S_{PP}$.

With Eq.~\eqref{eq:S_epsepsS_deltadelta}, we have derived the measurement imprecision of our detection scheme for the angular orientation of an anisotropic dipolar scatterer in a linearly polarized electromagnetic field. 
Equation~\eqref{eq:S_epsepsS_deltadelta} shows that the imprecision noise scales inversely with the number of photons scattered per unit time, which is given by $P_0/(\hbar\omega_0)$. This behavior is well known for any linear measurement limited by photon shot noise. Furthermore, the imprecision noise scales inversely with  the (square of the) optical anisotropy $\Delta$. Indeed, for an isotropic scatterer with $\Delta=0$, the measurement imprecision diverges since the scattered field cannot provide any information about the scatterer's orientation.

Finally, we note that in a realistic scenario, a dumbbell will possess at least some slight deviation from perfect cylindrical symmetry. In this case, the orientation vector $(\delta,\epsilon,\varphi)$ must be considered three dimensional with $\varphi$ the rotation angle around the long axis of the dumbbell. In App.~\ref{app:3D_anisotropic}, we show that to linear order in the orientation $(\delta,\epsilon,\varphi)$, the scattered field does not depend on the rotation $\varphi$. In addition, the correction of the detector signal, Eq.~\eqref{eq:Felix_P_det_delta}, due to the rotation $\varphi$ is suppressed by the factor $\Delta'/\Delta\ll1$, where $\Delta'$ characterizes the small anisotropy between the scatterer's short axes.
In conclusion, we can safely neglect any deviation from a perfect cylindrical symmetry due to its small effect on our result.

\section{Measurement backaction}
Having determined the imprecision of our measurement scheme for the orientation angles $\delta$ and $\epsilon$ of the anisotropic scatterer, we now turn to an analysis of the measurement backaction. This backaction arises as a torque noise driving the rotational motion of the scatterer and has been derived earlier~\cite{Stickler2016-decoherence,Zhong2017,Seberson2020}. 
To make this article self contained, we provide a particularly simple and didactic treatment here.

The system under consideration is still the anisotropic scatterer with polarizability $\tens{\alpha}=\alpha_0\text{diag}[1,1 - \Delta, 1- \Delta]$, aligned with its long axis to an $x$-polarized electric driving field, as outlined in Sec.~\ref{sec:system}.
The instantaneous torque $\vect\tau(t)$ experienced by a dipole moment $\vect{p}_\text{tot}(t)$ interacting with an electric field $\vect{E}_\text{tot}(t)$ is given by~\cite{Zangwill2013}
\begin{equation}
    \vect{\tau}(t) = \vect{p}_\text{tot}(t)\times\vect{E}_\text{tot}(t).
\end{equation}
Throughout this section, $\vect{p}_\text{tot}(t)$ and $\vect{E}_\text{tot}(t)$ are real valued, time dependent quantities. 
As before, all vectors are in the space-fixed Cartesian frame.
We split the electric field into the deterministic driving field $\vect{E}(t)=E_0\cos(\omega_0t)\nx$, and a fluctuating background field $\tilde{\vect{E}}(t)=[\tilde E_x(t),\tilde E_y(t), \tilde E_z(t)]$, such that the total field reads $\vect{E}_\text{tot}(t)=\vect{E}(t)+\tilde{\vect{E}}(t)$.
Throughout this work, we denote fluctuating quantities with a tilde and assume them to be statistically stationary, random variables of zero-mean. 
The total dipole moment of the scatterer is $\vect{p}_\text{tot}(t)=\tens\alpha\vect{E}_\text{tot}(t)$,
where we approximated the polarizability $\tens{\alpha}$ as purely real and given by its value at the optical frequency $\omega_0$.

We are interested in the fluctuating optical torque $\vect {\tilde \tau}(t)$, which to first order in the field fluctuations reads
\begin{equation}\label{eq:tau_firstOrder2}
\begin{split}
    \vect{\tilde \tau}(t) &= \alpha_0\vect E(t)\times \vect{\tilde E}(t) + [\tens{\alpha} \vect{\tilde E}(t)] \times \vect E(t) \\
    &=\alpha_0E_0 \Delta \cos(\omega_0 t) \colvec[0]{-\tilde E_z(t)}{\tilde E_y(t)}.
\end{split}
\end{equation}
This expression was derived at the equilibrium position $\delta= \epsilon=0$. The next order correction in $\delta, \epsilon$ is much smaller and can be neglected.
We thus find for the PSD of the $y$ component of the torque fluctuations $\tilde\tau_y$
\begin{equation}\label{eq:stautau_from_SEE}
\begin{split}
    S_{\tau \tau}^{yy}(\omega) &= \frac{1}{2\pi} \int_{-\infty}^\infty \diff t'~ \braket{\tilde \tau_y(t+t') \tilde \tau_y(t)} \expu^{\imu \omega t'} \\
    &= \frac{\alpha_0^2E_0^2}{4} \Delta^2 \left[ S_{EE}^{zz}(\omega + \omega_0) + S_{EE}^{zz}(\omega - \omega_0) \right], 
\end{split}
\end{equation}
where $S_{EE}^{zz}(\omega)$ is the PSD of $\tilde E_z(t)$ and the angle brackets denote ensemble and time average. An analogous expression is obtained for $S_{\tau\tau}^{zz}(\omega)$, which depends on  $S_{EE}^{yy}(\omega)$.
Note that in {Eq.~\eqref{eq:tau_firstOrder2}}, $\tilde{\vect{\tau}}(t)$ and $\tilde{\vect{E}}(t)$ could be both Hermitian quantum operators or classical, real-valued random processes with PSDs $S_{\tau\tau}^{ii}(\omega)$ and $S_{EE}^{ii}(\omega)$, $i\in\{x,y,z\}$, respectively.
In both cases, in the experimentally accessible regime of $|\omega|\ll\omega_0$, the PSDs of the torque fluctuations arise from the symmetric part of the field PSDs, which in vacuum are given by~\cite{[{This spectrum follows directly from Eq.~(C.31) on page~191 in }][{}]CohenTannoudji1989}
\begin{equation}\label{eq:S_EE}
\begin{split}
&\frac{1}{2}\left[S_{EE}^{ii}(\omega_0+\omega) +S_{EE}^{ii}(-\omega_0+\omega)\right] \\
\approx & ~ \frac{1}{2}\left[S_{EE}^{ii}(\omega_0) + S_{EE}^{ii}(-\omega_0)\right] = \frac{\hbar |\omega_0|^3}{12\pi^2 \epsilon_0 c^3}.
\end{split}
\end{equation}
In quantum theory, these field fluctuations can be seen as a consequence of the fact that the ladder operators describing the electromagnetic field do not commute.
In the framework of stochastic electrodynamics, these correlations are postulated to originate from a classical electromagnetic background field chosen such that each mode at frequency $\omega$ carries energy $\hbar\omega/2$~\cite{Marshall1963, Boyer1975}.
With these correlations for the fluctuating fields, the scattered power ${P_0=\omega_0^4\alpha_0^2E_0^2/(12\pi\epsilon_0c^3)}$, and the assumption $|\omega|\ll\omega_0$, the PSDs of the fluctuating torque components $\tilde\tau_y$ and $\tilde\tau_z$ read
\begin{equation}\label{eq:S_tautau_final2}
    S_{\tau \tau}^{yy} = S_{\tau \tau}^{zz} = \frac{1}{2\pi}   \Delta^2 \hbar^2\frac{P_0}{\hbar\omega_0}.
\end{equation}
The $x$ component of the torque fluctuations vanishes, according to Eq.~\eqref{eq:tau_firstOrder2}, such that we have $S^{xx}_{\tau\tau}(\omega) = 0$.

Let us compare our result for the torque noise to the  literature. Seberson and Robicheux have derived the energy heating rate $\dot E$ for a librator due to photon shot noise~\cite{Seberson2020}. When converting our result for the torque noise to an energy heating rate according to  $\dot E=\pi S_{\tau\tau}/I$~\cite{Clerk2010, bowen2015quantum}, we recover the result given in Ref.~\cite{Seberson2020}. We note that the simple relationship between torque noise and energy heating rate assumes simple harmonic oscillator dynamics of the libration mode. The situation is more involved when the scatterer spins around its axis of symmetry, which gives rise to a coupling of the libration modes~\cite{Seberson2019,Bang2020}.

Let us interpret Eq.~\eqref{eq:S_tautau_final2}.
The measurement backaction is proportional to the total number of photons scattered per unit time, $P_0/(\hbar \omega_0)$, since the photons are the source of the recoil torque. Since each photon carries an angular momentum of $\hbar$, the measurement backaction (which scales with the torque variance) is proportional to $\hbar^2$. Finally, the (square of the) optical anisotropy $\Delta$ enters Eq.~\eqref{eq:S_tautau_final2}, since an optical field cannot exert any torque on a lossless scatterer with vanishing anisotropy.

As a side, we note that our treatment provides an interesting insight into the role of classical noise, which may be present in the ($x$-polarized) driving laser. 
Such classical laser noise would increase $\tilde{ \vect E}$ beyond the level set by the vacuum fluctuations in Eq.~\eqref{eq:S_EE}.
Interestingly, while the (vacuum) fluctuations in the $y$ and $z$ components of the electric field dominate the torque fluctuations according to Eq.~\eqref{eq:tau_firstOrder2}, driving laser fluctuations $\tilde E_x$ make no appearance, such that classical laser noise does not lead to torque noise to linear order. For this reason, optomechanical systems based on levitated dipolar scatterers are particularly resilient to classical laser intensity noise. Classical noise that does indeed enter Eq.~\eqref{eq:tau_firstOrder2} is polarization noise. In contrast to classical intensity noise, polarization noise is easier to combat in practice since it can be suppressed with purely passive optical components.

As a further side note, our treatment provides some fundamental insight on how to engineer the torque shot noise acting on the anisotropic scatterer.  Equation~\eqref{eq:tau_firstOrder2} shows that $\tilde\tau_y$ depends exclusively on the vacuum fluctuations along the $z$ direction $\tilde E_z$ at the origin. To understand how to gain full control over that field component, it is instructive to consider the field impinging onto the scatterer in an expansion into vector spherical harmonics~\cite{Barrera1985, Carrascal1991}. In this basis, the only mode leading to a finite electric field along $z$ at the origin is the dipole mode $\vect{u}^{(z)}$. Therefore, it is the noise in this mode only which heats the libration angle $\epsilon$. To modify most efficiently the shot noise heating of this libration mode using squeezing, one therefore must feed squeezed vacuum into the incoming $\vect{u}^{(z)}$ mode. 
The effect of squeezing any other mode, e.g., the trapping beam, is limited by its mode overlap with $\vect{u}^{(z)}$.  
In analogy, for efficient suppression of shot noise heating of the angle $\delta$, the dipolar mode $\vect{u}^{(y)}$ must be squeezed.

Returning to our problem of optimal orientation detection, let us review our findings so far. In Sec.~\ref{sec:idealMeasurement}, we have computed the measurement imprecision of our optimal measurement scheme for the angles $\delta$ and $\epsilon$, given by Eq.~\eqref{eq:S_epsepsS_deltadelta}. 
In the current Section, we have derived the measurement backaction Eq.~\eqref{eq:S_tautau_final2} experienced by the scatterer due to the interference of the strong driving field with the electromagnetic vacuum fluctuations.
Importantly, the product between the measurement imprecision and the backaction yields the Heisenberg uncertainty limit according to~\cite{bowen2015quantum}
\begin{equation}
    S^{zz}_{\tau\tau}\times S_{\delta\delta} = S^{yy}_{\tau\tau}\times S_{\epsilon\epsilon} = \frac{1}{4\pi^2}\frac{\hbar^2}{4}.
\end{equation}
Thus, the measurement scheme presented in Sec.~\ref{sec:idealMeasurement} indeed represents an optimal scheme to resolve the angular orientation of the scatterer.

\section{Realistic measurement scheme}\label{sec:realisticMeasurement}
\begin{figure}[t]
\includegraphics[width=\columnwidth]{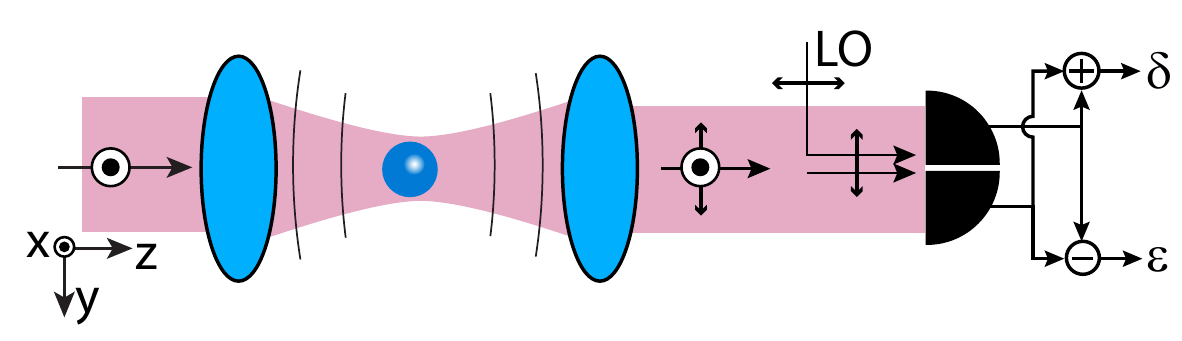}
\caption{Realistic detection scheme. A dumbbell (long axis along $x$) is trapped in an $x$-polarized field. The scattered field is collected in the forward direction and mixed with a strong $y$-polarized local oscillator (LO) before it is measured on a detector split in two halves along the $xz$ plane. The sum signal gives access to the orientation angle $\delta$, while the difference signal is proportional to the angle $\epsilon$ of the dumbbell.}
\label{fig:realisticScheme}
\end{figure}
The measurement scheme presented in Sec.~\ref{sec:idealMeasurement} has been proven to be ideal, albeit it is hardly practical. In this section, we analyze a simple and thus realistic detection scheme and quantify its efficiency. In our realistic scenario, illustrated in Fig.~\ref{fig:realisticScheme}, the scatterer is again trapped by an $x$-polarized beam propagating along the positive $z$ axis. The trapping beam is focused by a trapping lens and recollimated by a collection lens. These lenses also collect the scattered light.
According to Eq.~\eqref{eq:field_Dipole}, the information about the orientation angles $\delta$ and $\epsilon$ is contained in dipolar radiation modes which are symmetric with respect to the $xy$ plane. Therefore, detection in the forward and backward direction yields equal detection efficiency. 
Here, we choose to place our detector in the forward direction, but we note that the treatment of the scheme in backscattering is analogous. 
The field collimated by the collection lens is interfered with a strong $y$-polarized reference field serving as a local oscillator (LO). This LO field is assumed to be spatially homogeneous across the numerical aperture of the collection lens and in-phase with the signal field.

We stress that it is far from optimal to harness the transmitted trapping beam as a local oscillator, an observation made already in Ref.~\cite{Seberson2019}. Our objective is to determine the amplitude of the scattered field, which encodes the orientation angles $\delta$ and $\epsilon$. However, due to the Gouy phase shift in a focused field, the scattered field and the transmitted trapping beam are in different quadratures and therefore set up to detect the phase of the signal field, not its amplitude~\cite{Gittes1998}. 
The transmitted trapping beam thus only generates shot noise on the detector, without providing any information to linear order. When detecting in the forward direction, one should therefore either dump the trapping beam with a polarizing beamsplitter, or use an LO which is much stronger than the transmitted trapping beam, as we assume in our discussion and in Fig.~\ref{fig:realisticScheme}.

For mathematical convenience, we treat the interference of the LO field with the scattered field on the reference sphere of the collection lens. 
When collimated, the LO field has the polarization vector $\vect{n}_y = \sin(\phi)\vect{n}_\rho + \cos(\phi)\vect{n}_\phi$ in cylindrical coordinates $(\rho, \phi, z)$. According to the rules of field focusing,  as laid out in Chapter 3 of Ref.~\cite{Novotny2012}, on the lens' reference sphere this field takes the form
\begin{equation}\label{eq:referenceField_ypol}
    \vect{\mathcal E}_\text{lo} = \sqrt{\frac{P_\text{lo}}{\pi\text{NA}^2}\cos\theta}~[\sin\phi~ \ntheta + \cos\phi~\nphi]
\end{equation}
in spherical coordinates $(r,\theta,\phi)$.
We use the complex notation and normalization introduced in Sec.~\ref{sec:idealMeasurement}, such that integration of the LO field's modulus squared over the reference sphere within the numerical aperture $\text{NA}=\sin\theta_\mathrm{m}$ (given by the maximum collection angle $\theta_\mathrm{m}$)  yields the total power in the LO beam $P_\text{lo}$. 

To quantify the performance of our realistic measurement scheme for the angle $\delta$, we must analyze its imprecision noise $S_{\delta\delta}^\text{re}$ and compare it to that of the ideal measurement $S_{\delta\delta}$, given by Eq.~\eqref{eq:S_epsepsS_deltadelta}. The relevant figure of merit is thus the detection efficiency $\eta_\delta=S_{\delta\delta}/S_{\delta\delta}^\text{re}$. An analogous treatment must be carried out for the angle $\epsilon$.
As detailed in App.~\ref{app:ideal_ref_field}, the detection efficiencies associated with the measurements of $\delta$ and $\epsilon$ are given by the overlap integral between the chosen LO field Eq.~\eqref{eq:referenceField_ypol} and the signal field Eq.~\eqref{eq:field_Dipole}. According to Eq.~\eqref{eq:det_efficiency_general}, we find for the angle $\delta$ the detection efficiency
\begin{equation}\label{eq:eta_delta}
\begin{split}
    \eta_\delta &=\frac{1}{P_\text{lo}} \left( \int \diff \Omega~ \vect u^{(y)} \cdot \vect{ \mathcal E}_\text{lo} \right)^2= \frac{3U^2}{8\pi^2\text{NA}^2},
\end{split}
\end{equation}
with 
\begin{equation}\label{eq:U_theta_m}
\begin{aligned}
	U(\theta_\mathrm{m}) &= \int_0^{2\pi}\diff\phi\int_0^{\theta_\mathrm{m}} \diff\theta  \sin\theta \sqrt{\cos\theta} \\ & \qquad\times \left[\cos\theta\sin^2\phi + \cos^2\phi\right]
	\\ &= \frac{2\pi}{15}  \left(8-3\cos ^{\frac{5}{2}}(\theta_\mathrm{m}) - 5 \cos ^{\frac{3}{2}}(\theta_\mathrm{m})\right).
\end{aligned}
\end{equation}
\begin{figure}[t]
\includegraphics[width=\columnwidth]{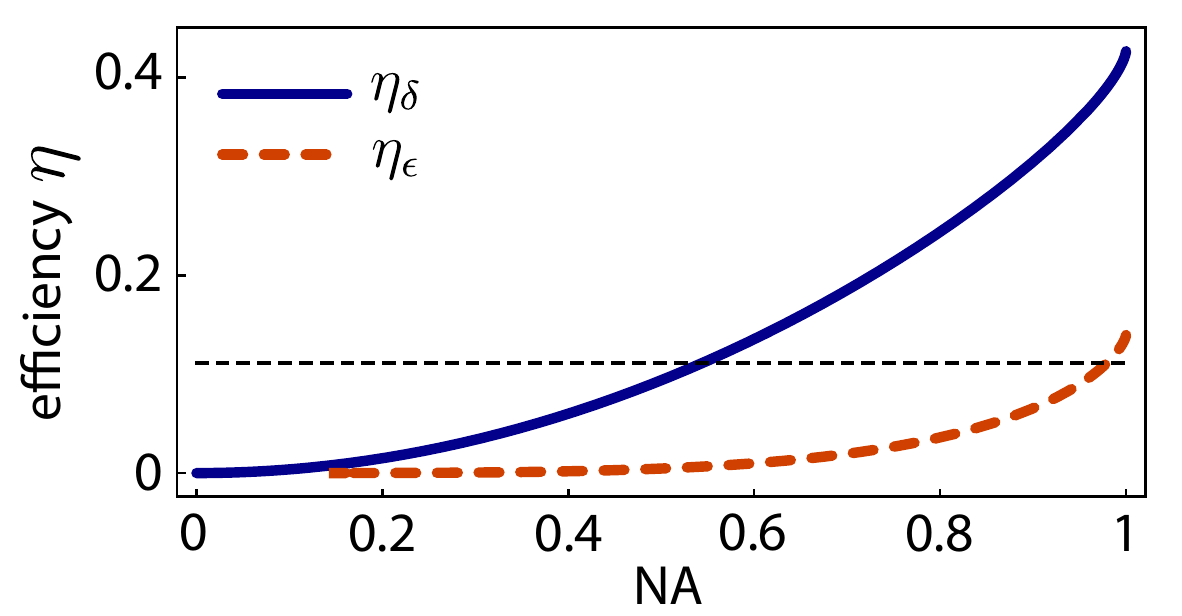}
\caption{
Detection efficiencies $\eta_\delta$ (blue solid line) and $\eta_\epsilon$ (orange dashed line) of our realistic detection scheme according to Eqs.~\eqref{eq:eta_delta} and \eqref{eq:eta_epsilon} as a function of the numerical aperture (NA) of the collection lens. The black dashed horizontal line marks $\eta=1/9$, the detection efficiency required to feedback cool to a phonon occupation number $n=1$.}
\label{fig:plotDetectionEfficiencies}
\end{figure}
In Fig.~\ref{fig:plotDetectionEfficiencies}, we plot the detection efficiency $\eta_\delta$ 
as a function of numerical aperture as the solid blue line.
To quantify the price we pay for the practicability of the realistic detection scheme, it is instructive to consider the detection efficiency for an NA of unity, where we find $\eta_{\delta}^\text{max} = 32/75$. This result is remarkably close to the value of one half, provided by ideal detection in a single half space. The deviation from the optimal result arises from the imperfect overlap of the LO field (a plane wave) with the signal field (radiation field of a dipole).

Having analyzed the detection efficiency for the angle $\delta$, let us turn to the angle $\epsilon$, which is encoded in the mode $\vect{u}^{(z)}$, whose field strictly points along $\ntheta$. When projected onto the local oscillator Eq.~\eqref{eq:referenceField_ypol}, the overlap integral contains the term $\sin\phi$, such that integration of $\phi$ over the domain $[0\ldots2\pi]$ yields a vanishing signal. 
Therefore, we split the detector into an ``upper'' half in the range $\phi=[0\ldots\pi]$, and a ``lower'' half corresponding to $\phi=[-\pi\ldots 0]$, see Fig.~\ref{fig:realisticScheme}. While summing the signals from these two halves leads to the detection of the angle $\delta$, as discussed, subtracting the signals provides access to the angle $\epsilon$. 
We thus find for the detection efficiency of the angle $\epsilon$
\begin{equation}\label{eq:eta_epsilon}
	\eta_\epsilon = \frac{6V^2}{\pi^2 \text{NA}^2},
\end{equation} with
\begin{equation}\label{}
	V(\theta_\mathrm{m}) = \int_0^{\theta_\mathrm{m}}\diff\theta~\sin^2\theta \sqrt{\cos\theta}.
\end{equation}
We plot $\eta_\epsilon$ as the orange dashed line in Fig.~\ref{fig:plotDetectionEfficiencies}. We note that $\eta_\epsilon$ is consistently lower than $\eta_\delta$. This observation can be explained by two reasons. First, the angle $\epsilon$ gives rise to a dipole moment induced in the scatterer pointing along $z$. This dipole radiates predominantly into the $xy$ plane, such that the signal is poorly captured by a collection optic of finite numerical aperture positioned along the $z$ axis. Second, even at unity numerical aperture, the overlap of the $y$-polarized LO field (or any other linearly polarized LO field) with the radiation mode $\vect{u}^{(z)}$ of the $z$ dipole is rather poor. 

It is possible to double the detection efficiency for $\epsilon$ in the forward direction by splitting the field at a polarizing beamsplitter into the $y$ component (and interfering it with a $y$-polarized local oscillator as just discussed), and the $x$ component, which has to be interfered with an $x$-polarized local oscillator. Due to the symmetry of the problem, the detector for the $x$ component should be split into a ``left'' and a ``right'' half. As already mentioned, a technical difficulty arises in forward scattering from the fact that the transmitted trapping field is $x$-polarized, such that the local oscillator in that polarization direction has to be much stronger than the trapping field. In backscattering, this technical problem is absent.

Let us discuss the consequences of our findings for measurement-based quantum control of libration modes of levitated nanoparticles. 
While a dumbbell spinning around its symmetry axis shows intricate dynamics, the libration modes reduce to two uncoupled harmonic oscillators in the absence of a spinning motion~\cite{Seberson2019}. Here, we analyze the cooling performance that can be expected from linear feedback control for harmonic-oscillator degrees of freedom, i.e., a librating particle that does not spin around its axis of symmetry.
For any harmonic oscillator, in a regime where the dominant heating rate is due to measurement backaction, linear feedback control allows for cooling to a mean phonon occupation $n=(\eta^{-1/2}-1)/2$, only limited by the detection efficiency $\eta$~\cite{Rossi2018}. Accordingly, cooling to unity phonon occupation (regarded as ``ground-state cooling'' by the community) requires a detection efficiency of $1/9$, which is marked as the black dashed horizontal line in Fig.~\ref{fig:plotDetectionEfficiencies}.  
At a realistically achievable numerical aperture of 0.9, we find for the angle $\epsilon$ a detection efficiency $\eta_\epsilon=0.07$ and an associated phonon occupation $n_\epsilon=1.4$. 
For the angle $\delta$, we find $\eta_\delta=0.31$, allowing for cooling to a phonon occupation $n_\delta=0.4$. Therefore, cooling the $\delta$ libration mode to its quantum ground state appears feasible with our realistic detection scheme. 
One should keep mind that our estimation is rather conservative given that the detection efficiencies plotted in Fig.~\eqref{fig:plotDetectionEfficiencies} can be significantly improved with some additional technical overhead. The efficiency $\eta_\delta$ can be doubled by implementing both a forward and a backward detection system. Furthermore, the efficiency $\eta_\epsilon$ plotted in Fig.~\eqref{fig:plotDetectionEfficiencies} can be quadrupled by implementing both a forward and a backward detection system and analyzing both polarization components.
Therefore, regarding the performance of a feasible detection scheme, ground-state cooling of librational motion appears within experimental reach.

\section{Conclusion}
In conclusion, we have theoretically analyzed the problem of measuring the orientation of an asymmetric scatterer with cylindrical symmetry in a linearly polarized light field. 
We have devised an optimal detection scheme 
for each of the two angles describing the scatterer's orientation, and we have analyzed the associated measurement imprecision. To prove that our measurement scheme is optimal, we have derived the associated measurement backaction and demonstrated that the imprecision-backaction product meets the Heisenberg uncertainty limit. Furthermore, we have proposed and analyzed a realistic detection scheme. The efficiency of our realistic measurement scheme is sufficient to allow for measurement-based feedback cooling of one angular degree of freedom of an anisotropic optically levitated particle to its quantum ground state of motion in the absence of spinning motion around the symmetry axis. With some additional technical overhead, ground-state cooling of both angular degrees of freedom is within reach.

\begin{acknowledgments}
We acknowledge financial support by ETH Grant ETH-47~20-2.
A.N.\ thanks the Jane and Aatos Erkko Foundation (Finland) for funding.
\end{acknowledgments}

\appendix

\section{Generalized linear measurement}\label{app:ideal_ref_field}
In this Appendix, we treat the case of a generalized linear measurement. 
First, we determine the optimal local oscillator (LO) field to extract a scalar quantity linearly encoded in a signal field.
Second, we derive the measurement imprecision of such a generalized linear measurement in the presence of photon shot noise in the case of an optimal and a sub-optimal LO field.

Suppose that a dimensionless quantity $x$ is linearly encoded in a signal field given by $x \vect{ \mathcal E}_s (\Og)$. This quantity $x$ could, for example, be one of the angles $\delta$ or $\epsilon$ in the main text. Assume that this signal field strikes a detector surface $A(\Omega)$, which is parametrized by $\Og$, at normal incidence.
A LO field $\vect {\mathcal E}_\text{lo}(\Og)$ is superposed with the signal field to interfere on the detector. Let this LO field carry a total power $P_\text{lo}=\int_A\diff\Omega~\vect{\mathcal E}_\text{lo}^*\cdot\vect{\mathcal E}_\text{lo}$, where the asterisk denotes complex conjugation.
Note that throughout this Appendix, as in Secs.~\ref{sec:idealMeasurement} and~\ref{sec:realisticMeasurement}, all fields are expressed in complex notation and normalized such that integration of their modulus squared over solid angle yields their power.
The detector signal fluctuates due to photon shot noise.
We now show that, in order to minimize the measurement imprecision for $x$, the field distribution $\vect {\mathcal E}_\text{lo}(\Og)$ must be in the same mode as the signal field $\vect{ \mathcal E}_s (\Og)$. 

We start with the power $P$ received by our detector
\begin{equation}\label{eq:detector_power}
\begin{split}
    P &= \int_A \diff \Og ~ |\vect{ \mathcal E}_\text{lo} + x\vect{ \mathcal E}_s|^2  = P_\text{lo} + C x,
\end{split}
\end{equation}
where we expanded to first order in the signal field 
and introduced the calibration factor $C$, which converts our signal from units of $x$ to units of power
\begin{equation}\label{eq:C_factor}
    C = 2\int_A \diff \Og~\re{ \vect{ \mathcal E}_\text{lo}^* \cdot \vect{ \mathcal E}_s }.
\end{equation}
The detected power $P$ has a constant offset given by the local oscillator power $P_\text{lo}$ and an interference term $Cx$, and therefore represents a linear measurement of the quantity $x$. 
The power fluctuations due to photon shot noise of the dominating power $P_\text{lo}$ have a white noise floor with power spectral density (PSD)~\cite{bowen2015quantum}
\begin{equation}\label{eq:app:SPP}
    S_{PP} = \frac{\hbar \omega_0}{2\pi}  P_\text{lo},
\end{equation}
such that the PSD of the imprecision noise of $x$ is 
\begin{equation}
    S_{xx} = \frac{S_{PP}}{C^2}. 
\end{equation}
Equation~\eqref{eq:app:SPP} arises in a rigorous quantum mechanical treatment by interpreting the (positive frequency part of the) field, $\vect {\mathcal E}$, as a bosonic annihilation operator. 
This operator is associated to a single electric-field mode characterized by the detector surface $A(\Omega)$.
For a coherent field, as in our case, 
we can write $\vect {\mathcal E}=\vect{\mathcal{\bar E}}+
\vect {\mathcal E}^{(+)}$ as a sum of a coherent amplitude $\vect{\mathcal{\bar E}}$ and the annihilation operator (or positive frequency part of the field)
$\vect {\mathcal E}^{(+)}$ accounting for the vacuum fluctuations.
The coherent amplitude $\vect{\mathcal{\bar E}}$ is identical to the amplitude of our treatment above and in the main text.
The power $P$ on our detector is then given by the sum of its mean value $\bar P$ [which is identical to our Eq.~\eqref{eq:detector_power}] and the operator $\hat P$, representing the fluctuations. To linear order in the field fluctuations, we find 
$\hat P = \int\diff\Omega ~ \left(\vect{ \mathcal{\bar E}}_\text{lo}^* \vect {\mathcal{E}}^{(+)}+\vect{ \mathcal{\bar E}}_\text{lo} {\vect {\mathcal E}}^{(-)} \right)$, where ${\vect {\mathcal E}}^{(-)}$ is the creation operator (or negative frequency part) of the vacuum field.
Finally, the two-time correlation of the detector power is computed as~\cite{Glauber1963, Carmichael1987}
\begin{equation}\label{eq:G_PP}
\begin{split}
    G_{PP}(\tau) &= \hbar\omega_0 \bar P\delta(\tau) + \braket{:\hat P(t+\tau)\hat{P}(t):} \\
    &= \hbar\omega_0 P_\text{lo}\delta(\tau) + \mathcal{O}(x).
\end{split}
\end{equation}
Here, $:~:$ denotes normal ordering of the bosonic operators, that is, all creation operators $\vect {\mathcal{E}}^{(-)}$ are moved before all annihilation operators $\vect {\mathcal{E}}^{(+)}$. 
With the electric field in the vacuum state $\ket{0}$, we have $\vect {\mathcal E}^{(+)}\ket{0} = 0$ and $\bra{0} \vect {\mathcal E}^{(-)} = 0$, and thus the normally ordered correlation term vanishes. 
The PSD $S_{PP}$ of Eq.~\eqref{eq:app:SPP} follows from Eq.~\eqref{eq:G_PP} by Fourier transformation~\footnotemark[1].

Let us return to our analysis of Eq.~\eqref{eq:C_factor}.
We now express both the LO field $\vect { \mathcal E}_\text{lo}=\sqrt{P_\text{lo}} \vect u_\text{lo}$ and the signal field $\vect { \mathcal E}_s=\sqrt{P_s} \vect u_s$ with their respective mode functions $\vect u_\text{lo}$ and $\vect u_s$, which are normalized according to
\begin{equation}\label{eq:ref_mode_norm_constr}
    \int_A \diff \Og ~ \vect u_\text{lo}^* \cdot \vect u_\text{lo} = \int_A \diff \Og ~ \vect u_s^* \cdot \vect u_s= 1.
\end{equation}
With the power of the signal field $P_s = \int_A\diff\Omega~\vect{ \mathcal E}_s^* \cdot \vect{ \mathcal E}_s$, we can express the imprecision PSD as an overlap integral between the LO and signal mode functions as
\begin{equation}\label{eq:Sxx_some_ref_mode}
    S_{xx} = \frac{\hbar\omega_0}{8\pi} \frac{1}{P_s} \left(\int_A\diff\Og~ \re{ \vect u_\text{lo}^*  \cdot \vect u_s } \right)^{-2}.
\end{equation}

Our task is now to find a mode function $\vect u_\text{lo}$ which minimizes $S_{xx}$ by maximising the mode overlap with $\vect u_s$. According to the Cauchy-Schwarz inequality, this maximized overlap is achieved when the LO mode is equal to the signal mode $\vect u_\text{lo}^\text{ideal} = \vect u_s$. We thus find the minimal imprecision noise PSD
\begin{equation}\label{eq:app:S_xx_ideal}
    S_{xx}^\text{ideal} = \frac{\hbar\omega_0}{8\pi} \frac{1}{P_s}.
\end{equation}
Let us apply this general finding to the problem of orientation detection from the main text. According to Eq.~\eqref{eq:field_Dipole}, the signal power associated with each of the angles $\delta$ and $\epsilon$ is given by $P_s=P_0\Delta^2$, such that Eq.~\eqref{eq:app:S_xx_ideal} yields Eq.~\eqref{eq:S_epsepsS_deltadelta} from the main text.

For a realistic detection system characterized by the (non-ideal) LO mode $\vect u_\text{lo}$, the detection efficiency $\eta$, defined as the ratio between the ideal imprecision noise $S_{xx}^\text{ideal}$ and the actual imprecision noise $S_{xx}$ as given by Eq.~\eqref{eq:Sxx_some_ref_mode}, is hence
\begin{equation}\label{eq:det_efficiency_general}
    \eta = \frac{S_{xx}^\text{ideal}}{S_{xx}} = \left( \int_A\diff\Og~ \re{\vect u_\text{lo}^* \cdot \vect u_\text{lo}^\text{ideal} } \right)^2.
\end{equation}
The detection efficiency is therefore given by the overlap integral between the used LO field $\vect u_\text{lo}$ and the ideal LO field $\vect u_\text{lo}^\text{ideal}$. We use Eq.~\eqref{eq:det_efficiency_general} in the main text to determine the realistic detection efficiencies for $\delta$ and $\epsilon$ in Eqs.~\eqref{eq:eta_delta} and \eqref{eq:eta_epsilon}.

\section{Dumbbell with broken cylindrical symmetry}\label{app:3D_anisotropic}
In the main text, we have assumed a scatterer with perfect cylindrical symmetry. Here, we assume the more realistic scenario that the scatterer still has a clearly defined long axis but additionally possesses a slight asymmetry in the transverse directions. Accordingly, we write the scatterer's polarizability as $\tens{\alpha}=\alpha_0\text{diag}[1 , 1- \Delta+\Delta'/2, 1 - \Delta - \Delta'/2]$ in a frame aligned with the particle axes. 
Here, $\Delta'<\Delta$ is a small anisotropy between the scatterer's short axes.
We still assume the first component, $\alpha_0$, to largely exceed the other two such that the particle aligns with the light's polarization axis. In addition to the rotations $\delta$ (around $z$) and $\epsilon$ (around $-y$), we allow for an arbitrary rotation $\varphi$ around the $x$ axis. The induced dipole moment $\vect p$ now reads
\begin{equation}
\begin{split}
    &\vect{p} = R_zR_{-y}R_x \tens \alpha R_{-x}R_{y}R_{-z}E_0 \vect n_x \\
    &= p_x \begin{pmatrix}
    1\\ \Delta\delta \\
    \Delta\epsilon 
    \end{pmatrix}
    - p_x \frac{\Delta'}{2} 
    \begin{pmatrix}
    0\\ \delta \cos{2\varphi} + \epsilon \sin{2\varphi} \\
    \delta \sin{2\varphi} - \epsilon \cos{2\varphi} 
    \end{pmatrix}\\
    &+ \mathcal{O}(\delta^2,\epsilon^2,\epsilon\delta),
\end{split}
\end{equation}
where $p_x=\alpha_0 E_0$ as in the main text. We expanded the expression to first order in $\delta$ and $\epsilon$, while $\varphi$ is not restricted. 
The first term in the final expression is the one discussed in the main text and due to the main anisotropy $\Delta$ between the main ($x$) axis and the other axes. 
In addition, we now find a second term due to the small anisotropy $\Delta'$. 
Its size is a factor $\Delta'/\Delta$ smaller than the main term. 
In this {second} term, the angles $\delta$ and $\epsilon$ appear in both the $y$ and $z$ component of the induced dipole moment. This leads to a mixing of the signals on the detector, depending on the rotation $\varphi$. 
Interestingly, the angle $\varphi$ only appears to second order (as a product with $\delta$ or $\epsilon$). A linear measurement of $\varphi$ is hence not possible.

\bibliography{bibliography}

\begin{thebibliography}{41}%
\makeatletter
\providecommand \@ifxundefined [1]{%
 \@ifx{#1\undefined}
}%
\providecommand \@ifnum [1]{%
 \ifnum #1\expandafter \@firstoftwo
 \else \expandafter \@secondoftwo
 \fi
}%
\providecommand \@ifx [1]{%
 \ifx #1\expandafter \@firstoftwo
 \else \expandafter \@secondoftwo
 \fi
}%
\providecommand \natexlab [1]{#1}%
\providecommand \enquote  [1]{``#1''}%
\providecommand \bibnamefont  [1]{#1}%
\providecommand \bibfnamefont [1]{#1}%
\providecommand \citenamefont [1]{#1}%
\providecommand \href@noop [0]{\@secondoftwo}%
\providecommand \href [0]{\begingroup \@sanitize@url \@href}%
\providecommand \@href[1]{\@@startlink{#1}\@@href}%
\providecommand \@@href[1]{\endgroup#1\@@endlink}%
\providecommand \@sanitize@url [0]{\catcode `\\12\catcode `\$12\catcode
  `\&12\catcode `\#12\catcode `\^12\catcode `\_12\catcode `\%12\relax}%
\providecommand \@@startlink[1]{}%
\providecommand \@@endlink[0]{}%
\providecommand \url  [0]{\begingroup\@sanitize@url \@url }%
\providecommand \@url [1]{\endgroup\@href {#1}{\urlprefix }}%
\providecommand \urlprefix  [0]{URL }%
\providecommand \Eprint [0]{\href }%
\providecommand \doibase [0]{http://dx.doi.org/}%
\providecommand \selectlanguage [0]{\@gobble}%
\providecommand \bibinfo  [0]{\@secondoftwo}%
\providecommand \bibfield  [0]{\@secondoftwo}%
\providecommand \translation [1]{[#1]}%
\providecommand \BibitemOpen [0]{}%
\providecommand \bibitemStop [0]{}%
\providecommand \bibitemNoStop [0]{.\EOS\space}%
\providecommand \EOS [0]{\spacefactor3000\relax}%
\providecommand \BibitemShut  [1]{\csname bibitem#1\endcsname}%
\let\auto@bib@innerbib\@empty
\bibitem [{\citenamefont {Aspelmeyer}\ \emph {et~al.}(2014)\citenamefont
  {Aspelmeyer}, \citenamefont {Kippenberg},\ and\ \citenamefont
  {Marquardt}}]{aspelmeyer2014cavity}%
  \BibitemOpen
  \bibfield  {author} {\bibinfo {author} {\bibfnamefont {M.}~\bibnamefont
  {Aspelmeyer}}, \bibinfo {author} {\bibfnamefont {T.}~\bibnamefont
  {Kippenberg}}, \ and\ \bibinfo {author} {\bibfnamefont {F.}~\bibnamefont
  {Marquardt}},\ }\href {https://books.google.ch/books?id=FG71AwAAQBAJ} {\emph
  {\bibinfo {title} {Cavity Optomechanics: Nano- and Micromechanical Resonators
  Interacting with Light}}},\ Quantum Science and Technology\ (\bibinfo
  {publisher} {Springer Berlin Heidelberg},\ \bibinfo {year}
  {2014})\BibitemShut {NoStop}%
\bibitem [{\citenamefont {Abbott}\ \emph {et~al.}(2016)\citenamefont {Abbott}
  \emph {et~al.}}]{Abbott2016}%
  \BibitemOpen
  \bibfield  {author} {\bibinfo {author} {\bibfnamefont {B.~P.}\ \bibnamefont
  {Abbott}} \emph {et~al.} (\bibinfo {collaboration} {LIGO Scientific
  Collaboration and Virgo Collaboration}),\ }\href {\doibase
  10.1103/PhysRevLett.116.061102} {\bibfield  {journal} {\bibinfo  {journal}
  {Phys. Rev. Lett.}\ }\textbf {\bibinfo {volume} {116}},\ \bibinfo {pages}
  {061102} (\bibinfo {year} {2016})}\BibitemShut {NoStop}%
\bibitem [{\citenamefont {Rossi}\ \emph {et~al.}(2018)\citenamefont {Rossi},
  \citenamefont {Mason}, \citenamefont {Chen}, \citenamefont {Tsaturyan},\ and\
  \citenamefont {Schliesser}}]{Rossi2018}%
  \BibitemOpen
  \bibfield  {author} {\bibinfo {author} {\bibfnamefont {M.}~\bibnamefont
  {Rossi}}, \bibinfo {author} {\bibfnamefont {D.}~\bibnamefont {Mason}},
  \bibinfo {author} {\bibfnamefont {J.}~\bibnamefont {Chen}}, \bibinfo {author}
  {\bibfnamefont {Y.}~\bibnamefont {Tsaturyan}}, \ and\ \bibinfo {author}
  {\bibfnamefont {A.}~\bibnamefont {Schliesser}},\ }\href {\doibase
  10.1038/s41586-018-0643-8} {\bibfield  {journal} {\bibinfo  {journal}
  {Nature}\ }\textbf {\bibinfo {volume} {563}},\ \bibinfo {pages} {53–58}
  (\bibinfo {year} {2018})}\BibitemShut {NoStop}%
\bibitem [{\citenamefont {Millen}\ \emph {et~al.}(2020)\citenamefont {Millen},
  \citenamefont {Monteiro}, \citenamefont {Pettit},\ and\ \citenamefont
  {Vamivakas}}]{Millen2020}%
  \BibitemOpen
  \bibfield  {author} {\bibinfo {author} {\bibfnamefont {J.}~\bibnamefont
  {Millen}}, \bibinfo {author} {\bibfnamefont {T.~S.}\ \bibnamefont
  {Monteiro}}, \bibinfo {author} {\bibfnamefont {R.}~\bibnamefont {Pettit}}, \
  and\ \bibinfo {author} {\bibfnamefont {A.~N.}\ \bibnamefont {Vamivakas}},\
  }\href {\doibase 10.1088/1361-6633/ab6100} {\bibfield  {journal} {\bibinfo
  {journal} {Reports on Progress in Physics}\ }\textbf {\bibinfo {volume}
  {83}},\ \bibinfo {pages} {026401} (\bibinfo {year} {2020})}\BibitemShut
  {NoStop}%
\bibitem [{\citenamefont {Reimann}\ \emph {et~al.}(2018)\citenamefont
  {Reimann}, \citenamefont {Doderer}, \citenamefont {Hebestreit}, \citenamefont
  {Diehl}, \citenamefont {Frimmer}, \citenamefont {Windey}, \citenamefont
  {Tebbenjohanns},\ and\ \citenamefont {Novotny}}]{Reimann2018}%
  \BibitemOpen
  \bibfield  {author} {\bibinfo {author} {\bibfnamefont {R.}~\bibnamefont
  {Reimann}}, \bibinfo {author} {\bibfnamefont {M.}~\bibnamefont {Doderer}},
  \bibinfo {author} {\bibfnamefont {E.}~\bibnamefont {Hebestreit}}, \bibinfo
  {author} {\bibfnamefont {R.}~\bibnamefont {Diehl}}, \bibinfo {author}
  {\bibfnamefont {M.}~\bibnamefont {Frimmer}}, \bibinfo {author} {\bibfnamefont
  {D.}~\bibnamefont {Windey}}, \bibinfo {author} {\bibfnamefont
  {F.}~\bibnamefont {Tebbenjohanns}}, \ and\ \bibinfo {author} {\bibfnamefont
  {L.}~\bibnamefont {Novotny}},\ }\href {\doibase
  10.1103/PhysRevLett.121.033602} {\bibfield  {journal} {\bibinfo  {journal}
  {Phys. Rev. Lett.}\ }\textbf {\bibinfo {volume} {121}},\ \bibinfo {pages}
  {033602} (\bibinfo {year} {2018})}\BibitemShut {NoStop}%
\bibitem [{\citenamefont {Ahn}\ \emph {et~al.}(2018)\citenamefont {Ahn},
  \citenamefont {Xu}, \citenamefont {Bang}, \citenamefont {Deng}, \citenamefont
  {Hoang}, \citenamefont {Han}, \citenamefont {Ma},\ and\ \citenamefont
  {Li}}]{Ahn2018}%
  \BibitemOpen
  \bibfield  {author} {\bibinfo {author} {\bibfnamefont {J.}~\bibnamefont
  {Ahn}}, \bibinfo {author} {\bibfnamefont {Z.}~\bibnamefont {Xu}}, \bibinfo
  {author} {\bibfnamefont {J.}~\bibnamefont {Bang}}, \bibinfo {author}
  {\bibfnamefont {Y.~H.}\ \bibnamefont {Deng}}, \bibinfo {author}
  {\bibfnamefont {T.~M.}\ \bibnamefont {Hoang}}, \bibinfo {author}
  {\bibfnamefont {Q.}~\bibnamefont {Han}}, \bibinfo {author} {\bibfnamefont
  {R.~M.}\ \bibnamefont {Ma}}, \ and\ \bibinfo {author} {\bibfnamefont
  {T.}~\bibnamefont {Li}},\ }\href {\doibase 10.1103/PhysRevLett.121.033603}
  {\bibfield  {journal} {\bibinfo  {journal} {Physical Review Letters}\
  }\textbf {\bibinfo {volume} {121}},\ \bibinfo {pages} {033603} (\bibinfo
  {year} {2018})}\BibitemShut {NoStop}%
\bibitem [{\citenamefont {Ahn}\ \emph {et~al.}(2020)\citenamefont {Ahn},
  \citenamefont {Xu}, \citenamefont {Bang}, \citenamefont {Ju}, \citenamefont
  {Gao},\ and\ \citenamefont {Li}}]{Ahn2020}%
  \BibitemOpen
  \bibfield  {author} {\bibinfo {author} {\bibfnamefont {J.}~\bibnamefont
  {Ahn}}, \bibinfo {author} {\bibfnamefont {Z.}~\bibnamefont {Xu}}, \bibinfo
  {author} {\bibfnamefont {J.}~\bibnamefont {Bang}}, \bibinfo {author}
  {\bibfnamefont {P.}~\bibnamefont {Ju}}, \bibinfo {author} {\bibfnamefont
  {X.}~\bibnamefont {Gao}}, \ and\ \bibinfo {author} {\bibfnamefont
  {T.}~\bibnamefont {Li}},\ }\href {\doibase 10.1038/s41565-019-0605-9}
  {\bibfield  {journal} {\bibinfo  {journal} {Nature Nanotechnology}\ }\textbf
  {\bibinfo {volume} {15}},\ \bibinfo {pages} {89} (\bibinfo {year}
  {2020})}\BibitemShut {NoStop}%
\bibitem [{\citenamefont {Bang}\ \emph {et~al.}(2020)\citenamefont {Bang},
  \citenamefont {Seberson}, \citenamefont {Ju}, \citenamefont {Ahn},
  \citenamefont {Xu}, \citenamefont {Gao}, \citenamefont {Robicheaux},\ and\
  \citenamefont {Li}}]{Bang2020}%
  \BibitemOpen
  \bibfield  {author} {\bibinfo {author} {\bibfnamefont {J.}~\bibnamefont
  {Bang}}, \bibinfo {author} {\bibfnamefont {T.}~\bibnamefont {Seberson}},
  \bibinfo {author} {\bibfnamefont {P.}~\bibnamefont {Ju}}, \bibinfo {author}
  {\bibfnamefont {J.}~\bibnamefont {Ahn}}, \bibinfo {author} {\bibfnamefont
  {Z.}~\bibnamefont {Xu}}, \bibinfo {author} {\bibfnamefont {X.}~\bibnamefont
  {Gao}}, \bibinfo {author} {\bibfnamefont {F.}~\bibnamefont {Robicheaux}}, \
  and\ \bibinfo {author} {\bibfnamefont {T.}~\bibnamefont {Li}},\ }\href
  {\doibase 10.1103/PhysRevResearch.2.043054} {\bibfield  {journal} {\bibinfo
  {journal} {Phys. Rev. Research}\ }\textbf {\bibinfo {volume} {2}},\ \bibinfo
  {pages} {043054} (\bibinfo {year} {2020})}\BibitemShut {NoStop}%
\bibitem [{\citenamefont {Delord}\ \emph {et~al.}(2020)\citenamefont {Delord},
  \citenamefont {Huillery}, \citenamefont {Nicolas},\ and\ \citenamefont
  {H{\'e}tet}}]{Delord2020}%
  \BibitemOpen
  \bibfield  {author} {\bibinfo {author} {\bibfnamefont {T.}~\bibnamefont
  {Delord}}, \bibinfo {author} {\bibfnamefont {P.}~\bibnamefont {Huillery}},
  \bibinfo {author} {\bibfnamefont {L.}~\bibnamefont {Nicolas}}, \ and\
  \bibinfo {author} {\bibfnamefont {G.}~\bibnamefont {H{\'e}tet}},\ }\href
  {\doibase 10.1038/s41586-020-2133-z} {\bibfield  {journal} {\bibinfo
  {journal} {Nature}\ }\textbf {\bibinfo {volume} {580}},\ \bibinfo {pages}
  {56} (\bibinfo {year} {2020})}\BibitemShut {NoStop}%
\bibitem [{\citenamefont {Stickler}\ \emph {et~al.}(2021)\citenamefont
  {Stickler}, \citenamefont {Hornberger},\ and\ \citenamefont
  {Kim}}]{Stickler2021}%
  \BibitemOpen
  \bibfield  {author} {\bibinfo {author} {\bibfnamefont {B.~A.}\ \bibnamefont
  {Stickler}}, \bibinfo {author} {\bibfnamefont {K.}~\bibnamefont
  {Hornberger}}, \ and\ \bibinfo {author} {\bibfnamefont {M.~S.}\ \bibnamefont
  {Kim}},\ }\href {\doibase 10.1038/s42254-021-00335-0} {\bibfield  {journal}
  {\bibinfo  {journal} {Nature Reviews Physics}\ } (\bibinfo {year} {2021}),\
  10.1038/s42254-021-00335-0}\BibitemShut {NoStop}%
\bibitem [{\citenamefont {Sch\"afer}\ \emph {et~al.}(2021)\citenamefont
  {Sch\"afer}, \citenamefont {Rudolph}, \citenamefont {Hornberger},\ and\
  \citenamefont {Stickler}}]{Schaefer2021}%
  \BibitemOpen
  \bibfield  {author} {\bibinfo {author} {\bibfnamefont {J.}~\bibnamefont
  {Sch\"afer}}, \bibinfo {author} {\bibfnamefont {H.}~\bibnamefont {Rudolph}},
  \bibinfo {author} {\bibfnamefont {K.}~\bibnamefont {Hornberger}}, \ and\
  \bibinfo {author} {\bibfnamefont {B.~A.}\ \bibnamefont {Stickler}},\ }\href
  {\doibase 10.1103/PhysRevLett.126.163603} {\bibfield  {journal} {\bibinfo
  {journal} {Phys. Rev. Lett.}\ }\textbf {\bibinfo {volume} {126}},\ \bibinfo
  {pages} {163603} (\bibinfo {year} {2021})}\BibitemShut {NoStop}%
\bibitem [{\citenamefont {Rudolph}\ \emph {et~al.}(2021)\citenamefont
  {Rudolph}, \citenamefont {Sch\"afer}, \citenamefont {Stickler},\ and\
  \citenamefont {Hornberger}}]{Rudolph2021}%
  \BibitemOpen
  \bibfield  {author} {\bibinfo {author} {\bibfnamefont {H.}~\bibnamefont
  {Rudolph}}, \bibinfo {author} {\bibfnamefont {J.}~\bibnamefont {Sch\"afer}},
  \bibinfo {author} {\bibfnamefont {B.~A.}\ \bibnamefont {Stickler}}, \ and\
  \bibinfo {author} {\bibfnamefont {K.}~\bibnamefont {Hornberger}},\ }\href
  {\doibase 10.1103/PhysRevA.103.043514} {\bibfield  {journal} {\bibinfo
  {journal} {Phys. Rev. A}\ }\textbf {\bibinfo {volume} {103}},\ \bibinfo
  {pages} {043514} (\bibinfo {year} {2021})}\BibitemShut {NoStop}%
\bibitem [{\citenamefont {Zhong}\ and\ \citenamefont
  {Robicheaux}(2017)}]{Zhong2017}%
  \BibitemOpen
  \bibfield  {author} {\bibinfo {author} {\bibfnamefont {C.}~\bibnamefont
  {Zhong}}\ and\ \bibinfo {author} {\bibfnamefont {F.}~\bibnamefont
  {Robicheaux}},\ }\href {\doibase 10.1103/PhysRevA.95.053421} {\bibfield
  {journal} {\bibinfo  {journal} {Phys. Rev. A}\ }\textbf {\bibinfo {volume}
  {95}},\ \bibinfo {pages} {053421} (\bibinfo {year} {2017})}\BibitemShut
  {NoStop}%
\bibitem [{\citenamefont {Seberson}\ and\ \citenamefont
  {Robicheaux}(2019)}]{Seberson2019}%
  \BibitemOpen
  \bibfield  {author} {\bibinfo {author} {\bibfnamefont {T.}~\bibnamefont
  {Seberson}}\ and\ \bibinfo {author} {\bibfnamefont {F.}~\bibnamefont
  {Robicheaux}},\ }\href {\doibase 10.1103/PhysRevA.99.013821} {\bibfield
  {journal} {\bibinfo  {journal} {Physical Review A}\ }\textbf {\bibinfo
  {volume} {99}},\ \bibinfo {pages} {013821} (\bibinfo {year} {2019})},\
  \Eprint {http://arxiv.org/abs/1810.01797} {1810.01797} \BibitemShut {NoStop}%
\bibitem [{\citenamefont {Stickler}\ \emph
  {et~al.}(2016{\natexlab{a}})\citenamefont {Stickler}, \citenamefont
  {Nimmrichter}, \citenamefont {Martinetz}, \citenamefont {Kuhn}, \citenamefont
  {Arndt},\ and\ \citenamefont {Hornberger}}]{Stickler2016}%
  \BibitemOpen
  \bibfield  {author} {\bibinfo {author} {\bibfnamefont {B.~A.}\ \bibnamefont
  {Stickler}}, \bibinfo {author} {\bibfnamefont {S.}~\bibnamefont
  {Nimmrichter}}, \bibinfo {author} {\bibfnamefont {L.}~\bibnamefont
  {Martinetz}}, \bibinfo {author} {\bibfnamefont {S.}~\bibnamefont {Kuhn}},
  \bibinfo {author} {\bibfnamefont {M.}~\bibnamefont {Arndt}}, \ and\ \bibinfo
  {author} {\bibfnamefont {K.}~\bibnamefont {Hornberger}},\ }\href {\doibase
  10.1103/PhysRevA.94.033818} {\bibfield  {journal} {\bibinfo  {journal} {Phys.
  Rev. A}\ }\textbf {\bibinfo {volume} {94}},\ \bibinfo {pages} {033818}
  (\bibinfo {year} {2016}{\natexlab{a}})}\BibitemShut {NoStop}%
\bibitem [{\citenamefont {Stickler}\ \emph
  {et~al.}(2018{\natexlab{a}})\citenamefont {Stickler}, \citenamefont
  {Schrinski},\ and\ \citenamefont {Hornberger}}]{Stickler2018}%
  \BibitemOpen
  \bibfield  {author} {\bibinfo {author} {\bibfnamefont {B.~A.}\ \bibnamefont
  {Stickler}}, \bibinfo {author} {\bibfnamefont {B.}~\bibnamefont {Schrinski}},
  \ and\ \bibinfo {author} {\bibfnamefont {K.}~\bibnamefont {Hornberger}},\
  }\href {\doibase 10.1103/PhysRevLett.121.040401} {\bibfield  {journal}
  {\bibinfo  {journal} {Phys. Rev. Lett.}\ }\textbf {\bibinfo {volume} {121}},\
  \bibinfo {pages} {040401} (\bibinfo {year} {2018}{\natexlab{a}})}\BibitemShut
  {NoStop}%
\bibitem [{\citenamefont {Stickler}\ \emph
  {et~al.}(2018{\natexlab{b}})\citenamefont {Stickler}, \citenamefont
  {Papendell}, \citenamefont {Kuhn}, \citenamefont {Schrinski}, \citenamefont
  {Millen}, \citenamefont {Arndt},\ and\ \citenamefont
  {Hornberger}}]{Stickler2018rotationquantum}%
  \BibitemOpen
  \bibfield  {author} {\bibinfo {author} {\bibfnamefont {B.~A.}\ \bibnamefont
  {Stickler}}, \bibinfo {author} {\bibfnamefont {B.}~\bibnamefont {Papendell}},
  \bibinfo {author} {\bibfnamefont {S.}~\bibnamefont {Kuhn}}, \bibinfo {author}
  {\bibfnamefont {B.}~\bibnamefont {Schrinski}}, \bibinfo {author}
  {\bibfnamefont {J.}~\bibnamefont {Millen}}, \bibinfo {author} {\bibfnamefont
  {M.}~\bibnamefont {Arndt}}, \ and\ \bibinfo {author} {\bibfnamefont
  {K.}~\bibnamefont {Hornberger}},\ }\href {\doibase 10.1088/1367-2630/aaece4}
  {\bibfield  {journal} {\bibinfo  {journal} {New J. Phys.}\ }\textbf {\bibinfo
  {volume} {20}},\ \bibinfo {pages} {122001} (\bibinfo {year}
  {2018}{\natexlab{b}})}\BibitemShut {NoStop}%
\bibitem [{\citenamefont {Romero-Isart}(2017)}]{Romero-Isart2017}%
  \BibitemOpen
  \bibfield  {author} {\bibinfo {author} {\bibfnamefont {O.}~\bibnamefont
  {Romero-Isart}},\ }\href {\doibase 10.1088/1367-2630/aa99bf} {\bibfield
  {journal} {\bibinfo  {journal} {New Journal of Physics}\ }\textbf {\bibinfo
  {volume} {19}},\ \bibinfo {pages} {123029} (\bibinfo {year}
  {2017})}\BibitemShut {NoStop}%
\bibitem [{\citenamefont {Manjavacas}\ and\ \citenamefont {{Garc{\'{i}}a De
  Abajo}}(2010)}]{Manjavacas2010}%
  \BibitemOpen
  \bibfield  {author} {\bibinfo {author} {\bibfnamefont {A.}~\bibnamefont
  {Manjavacas}}\ and\ \bibinfo {author} {\bibfnamefont {F.~J.}\ \bibnamefont
  {{Garc{\'{i}}a De Abajo}}},\ }\href {\doibase 10.1103/PhysRevLett.105.113601}
  {\bibfield  {journal} {\bibinfo  {journal} {Phys. Rev. Lett.}\ }\textbf
  {\bibinfo {volume} {105}},\ \bibinfo {pages} {113601} (\bibinfo {year}
  {2010})}\BibitemShut {NoStop}%
\bibitem [{\citenamefont {Zhao}\ \emph {et~al.}(2012)\citenamefont {Zhao},
  \citenamefont {Manjavacas}, \citenamefont {{Garc{\'{i}}a De Abajo}},\ and\
  \citenamefont {Pendry}}]{Zhao2012}%
  \BibitemOpen
  \bibfield  {author} {\bibinfo {author} {\bibfnamefont {R.}~\bibnamefont
  {Zhao}}, \bibinfo {author} {\bibfnamefont {A.}~\bibnamefont {Manjavacas}},
  \bibinfo {author} {\bibfnamefont {F.~J.}\ \bibnamefont {{Garc{\'{i}}a De
  Abajo}}}, \ and\ \bibinfo {author} {\bibfnamefont {J.~B.}\ \bibnamefont
  {Pendry}},\ }\href {\doibase 10.1103/PhysRevLett.109.123604} {\bibfield
  {journal} {\bibinfo  {journal} {Phys. Rev. Lett.}\ }\textbf {\bibinfo
  {volume} {109}},\ \bibinfo {pages} {123604} (\bibinfo {year}
  {2012})}\BibitemShut {NoStop}%
\bibitem [{\citenamefont {Xu}\ and\ \citenamefont {Li}(2017)}]{Xu2017}%
  \BibitemOpen
  \bibfield  {author} {\bibinfo {author} {\bibfnamefont {Z.}~\bibnamefont
  {Xu}}\ and\ \bibinfo {author} {\bibfnamefont {T.}~\bibnamefont {Li}},\ }\href
  {\doibase 10.1103/PhysRevA.96.033843} {\bibfield  {journal} {\bibinfo
  {journal} {Phys. Rev. A}\ }\textbf {\bibinfo {volume} {96}},\ \bibinfo
  {pages} {033843} (\bibinfo {year} {2017})}\BibitemShut {NoStop}%
\bibitem [{\citenamefont {Manjavacas}\ \emph {et~al.}(2017)\citenamefont
  {Manjavacas}, \citenamefont {Rodr{\'{i}}guez-Fortu{\~{n}}o}, \citenamefont
  {{Javier Garc{\'{i}}a De Abajo}},\ and\ \citenamefont
  {Zayats}}]{Manjavacas2017}%
  \BibitemOpen
  \bibfield  {author} {\bibinfo {author} {\bibfnamefont {A.}~\bibnamefont
  {Manjavacas}}, \bibinfo {author} {\bibfnamefont {F.~J.}\ \bibnamefont
  {Rodr{\'{i}}guez-Fortu{\~{n}}o}}, \bibinfo {author} {\bibfnamefont
  {F.}~\bibnamefont {{Javier Garc{\'{i}}a De Abajo}}}, \ and\ \bibinfo {author}
  {\bibfnamefont {A.~V.}\ \bibnamefont {Zayats}},\ }\href {\doibase
  10.1103/PhysRevLett.118.133605} {\bibfield  {journal} {\bibinfo  {journal}
  {Phys. Rev. Lett.}\ }\textbf {\bibinfo {volume} {118}},\ \bibinfo {pages}
  {133605} (\bibinfo {year} {2017})}\BibitemShut {NoStop}%
\bibitem [{\citenamefont {Magrini}\ \emph {et~al.}(2021)\citenamefont
  {Magrini}, \citenamefont {Rosenzweig}, \citenamefont {Bach}, \citenamefont
  {Deutschmann-Olek}, \citenamefont {Hofer}, \citenamefont {Hong},
  \citenamefont {Kiesel}, \citenamefont {Kugi},\ and\ \citenamefont
  {Aspelmeyer}}]{Magrini2021}%
  \BibitemOpen
  \bibfield  {author} {\bibinfo {author} {\bibfnamefont {L.}~\bibnamefont
  {Magrini}}, \bibinfo {author} {\bibfnamefont {P.}~\bibnamefont {Rosenzweig}},
  \bibinfo {author} {\bibfnamefont {C.}~\bibnamefont {Bach}}, \bibinfo {author}
  {\bibfnamefont {A.}~\bibnamefont {Deutschmann-Olek}}, \bibinfo {author}
  {\bibfnamefont {S.~G.}\ \bibnamefont {Hofer}}, \bibinfo {author}
  {\bibfnamefont {S.}~\bibnamefont {Hong}}, \bibinfo {author} {\bibfnamefont
  {N.}~\bibnamefont {Kiesel}}, \bibinfo {author} {\bibfnamefont
  {A.}~\bibnamefont {Kugi}}, \ and\ \bibinfo {author} {\bibfnamefont
  {M.}~\bibnamefont {Aspelmeyer}},\ }\href {\doibase
  10.1038/s41586-021-03602-3} {\bibfield  {journal} {\bibinfo  {journal}
  {Nature}\ }\textbf {\bibinfo {volume} {595}},\ \bibinfo {pages} {373}
  (\bibinfo {year} {2021})}\BibitemShut {NoStop}%
\bibitem [{\citenamefont {Tebbenjohanns}\ \emph {et~al.}(2021)\citenamefont
  {Tebbenjohanns}, \citenamefont {Mattana}, \citenamefont {Rossi},
  \citenamefont {Frimmer},\ and\ \citenamefont {Novotny}}]{Tebbenjohanns2021}%
  \BibitemOpen
  \bibfield  {author} {\bibinfo {author} {\bibfnamefont {F.}~\bibnamefont
  {Tebbenjohanns}}, \bibinfo {author} {\bibfnamefont {M.~L.}\ \bibnamefont
  {Mattana}}, \bibinfo {author} {\bibfnamefont {M.}~\bibnamefont {Rossi}},
  \bibinfo {author} {\bibfnamefont {M.}~\bibnamefont {Frimmer}}, \ and\
  \bibinfo {author} {\bibfnamefont {L.}~\bibnamefont {Novotny}},\ }\href
  {\doibase 10.1038/s41586-021-03617-w} {\bibfield  {journal} {\bibinfo
  {journal} {Nature}\ }\textbf {\bibinfo {volume} {595}},\ \bibinfo {pages}
  {378} (\bibinfo {year} {2021})}\BibitemShut {NoStop}%
\bibitem [{\citenamefont {Tebbenjohanns}\ \emph {et~al.}(2019)\citenamefont
  {Tebbenjohanns}, \citenamefont {Frimmer},\ and\ \citenamefont
  {Novotny}}]{Tebbenjohanns2019Efficiency}%
  \BibitemOpen
  \bibfield  {author} {\bibinfo {author} {\bibfnamefont {F.}~\bibnamefont
  {Tebbenjohanns}}, \bibinfo {author} {\bibfnamefont {M.}~\bibnamefont
  {Frimmer}}, \ and\ \bibinfo {author} {\bibfnamefont {L.}~\bibnamefont
  {Novotny}},\ }\href {\doibase 10.1103/PhysRevA.100.043821} {\bibfield
  {journal} {\bibinfo  {journal} {Phys. Rev. A}\ }\textbf {\bibinfo {volume}
  {100}},\ \bibinfo {pages} {043821} (\bibinfo {year} {2019})}\BibitemShut
  {NoStop}%
\bibitem [{\citenamefont {van~der Laan}\ \emph {et~al.}(2021)\citenamefont
  {van~der Laan}, \citenamefont {Tebbenjohanns}, \citenamefont {Reimann},
  \citenamefont {Vijayan}, \citenamefont {Novotny},\ and\ \citenamefont
  {Frimmer}}]{vanderLaan2020shotnoise}%
  \BibitemOpen
  \bibfield  {author} {\bibinfo {author} {\bibfnamefont {F.}~\bibnamefont
  {van~der Laan}}, \bibinfo {author} {\bibfnamefont {F.}~\bibnamefont
  {Tebbenjohanns}}, \bibinfo {author} {\bibfnamefont {R.}~\bibnamefont
  {Reimann}}, \bibinfo {author} {\bibfnamefont {J.}~\bibnamefont {Vijayan}},
  \bibinfo {author} {\bibfnamefont {L.}~\bibnamefont {Novotny}}, \ and\
  \bibinfo {author} {\bibfnamefont {M.}~\bibnamefont {Frimmer}},\ }\href
  {\doibase 10.1103/PhysRevLett.127.123605} {\bibfield  {journal} {\bibinfo
  {journal} {Phys. Rev. Lett.}\ }\textbf {\bibinfo {volume} {127}},\ \bibinfo
  {pages} {123605} (\bibinfo {year} {2021})}\BibitemShut {NoStop}%
\bibitem [{\citenamefont {Novotny}\ and\ \citenamefont
  {Hecht}(2012)}]{Novotny2012}%
  \BibitemOpen
  \bibfield  {author} {\bibinfo {author} {\bibfnamefont {L.}~\bibnamefont
  {Novotny}}\ and\ \bibinfo {author} {\bibfnamefont {B.}~\bibnamefont
  {Hecht}},\ }\href {\doibase 10.1017/CBO9780511813535} {\emph {\bibinfo
  {title} {Principles of Nano-Optics}}}\ (\bibinfo  {publisher} {Cambridge
  University Press},\ \bibinfo {year} {2012})\BibitemShut {NoStop}%
\bibitem [{\citenamefont {Bowen}\ and\ \citenamefont
  {Milburn}(2015)}]{bowen2015quantum}%
  \BibitemOpen
  \bibfield  {author} {\bibinfo {author} {\bibfnamefont {W.}~\bibnamefont
  {Bowen}}\ and\ \bibinfo {author} {\bibfnamefont {G.}~\bibnamefont
  {Milburn}},\ }\href {https://books.google.ch/books?id=xqlcrgEACAAJ} {\emph
  {\bibinfo {title} {Quantum Optomechanics}}}\ (\bibinfo  {publisher} {Taylor
  \& Francis},\ \bibinfo {year} {2015})\BibitemShut {NoStop}%
\bibitem [{Note1()}]{Note1}%
  \BibitemOpen
  \bibinfo {note} {\label {fn:psd_normalization}We normalize the PSD
  $S_{xx}(\omega )$ of a signal $x(t)$ such that $\mathinner {\langle
  {x(t)^2}\rangle }=\DOTSI \intop \ilimits@ _{-\infty }^\infty \protect \text
  {d}\omega ~S_{xx}(\omega )$.}\BibitemShut {Stop}%
\bibitem [{\citenamefont {Stickler}\ \emph
  {et~al.}(2016{\natexlab{b}})\citenamefont {Stickler}, \citenamefont
  {Papendell},\ and\ \citenamefont {Hornberger}}]{Stickler2016-decoherence}%
  \BibitemOpen
  \bibfield  {author} {\bibinfo {author} {\bibfnamefont {B.~A.}\ \bibnamefont
  {Stickler}}, \bibinfo {author} {\bibfnamefont {B.}~\bibnamefont {Papendell}},
  \ and\ \bibinfo {author} {\bibfnamefont {K.}~\bibnamefont {Hornberger}},\
  }\href {\doibase 10.1103/PhysRevA.94.033828} {\bibfield  {journal} {\bibinfo
  {journal} {Phys. Rev. A}\ }\textbf {\bibinfo {volume} {94}},\ \bibinfo
  {pages} {033828} (\bibinfo {year} {2016}{\natexlab{b}})}\BibitemShut
  {NoStop}%
\bibitem [{\citenamefont {Seberson}\ and\ \citenamefont
  {Robicheaux}(2020)}]{Seberson2020}%
  \BibitemOpen
  \bibfield  {author} {\bibinfo {author} {\bibfnamefont {T.}~\bibnamefont
  {Seberson}}\ and\ \bibinfo {author} {\bibfnamefont {F.}~\bibnamefont
  {Robicheaux}},\ }\href {\doibase 10.1103/PhysRevA.102.033505} {\bibfield
  {journal} {\bibinfo  {journal} {Physical Review A}\ }\textbf {\bibinfo
  {volume} {102}},\ \bibinfo {pages} {033505} (\bibinfo {year} {2020})},\
  \Eprint {http://arxiv.org/abs/1909.06469} {1909.06469} \BibitemShut {NoStop}%
\bibitem [{\citenamefont {Zangwill}(2013)}]{Zangwill2013}%
  \BibitemOpen
  \bibfield  {author} {\bibinfo {author} {\bibfnamefont {A.}~\bibnamefont
  {Zangwill}},\ }\href {https://books.google.ch/books?id=tEYSUegp9WYC} {\emph
  {\bibinfo {title} {Modern Electrodynamics}}},\ Modern Electrodynamics\
  (\bibinfo  {publisher} {Cambridge University Press},\ \bibinfo {year}
  {2013})\BibitemShut {NoStop}%
\bibitem [{\citenamefont {Cohen-Tannoudji}\ \emph {et~al.}(1989)\citenamefont
  {Cohen-Tannoudji}, \citenamefont {Dupont-Roc},\ and\ \citenamefont
  {Grynberg}}]{CohenTannoudji1989}%
  \BibitemOpen
  \bibfield  {author} {\bibinfo {author} {\bibfnamefont {C.}~\bibnamefont
  {Cohen-Tannoudji}}, \bibinfo {author} {\bibfnamefont {J.}~\bibnamefont
  {Dupont-Roc}}, \ and\ \bibinfo {author} {\bibfnamefont {G.}~\bibnamefont
  {Grynberg}},\ }\href@noop {} {\emph {\bibinfo {title} {Photons and Atoms -
  Introduction to Quantum Electrodynamics}}}\ (\bibinfo  {publisher} {Wiley},\
  \bibinfo {address} {New York},\ \bibinfo {year} {1989})\BibitemShut {NoStop}%
\bibitem [{\citenamefont {Marshall}(1963)}]{Marshall1963}%
  \BibitemOpen
  \bibfield  {author} {\bibinfo {author} {\bibfnamefont {T.~W.}\ \bibnamefont
  {Marshall}},\ }\href {\doibase 10.1098/rspa.1963.0220} {\bibfield  {journal}
  {\bibinfo  {journal} {Proceedings of the Royal Society of London. Series A.
  Mathematical and Physical Sciences}\ }\textbf {\bibinfo {volume} {276}},\
  \bibinfo {pages} {475} (\bibinfo {year} {1963})}\BibitemShut {NoStop}%
\bibitem [{\citenamefont {Boyer}(1975)}]{Boyer1975}%
  \BibitemOpen
  \bibfield  {author} {\bibinfo {author} {\bibfnamefont {T.~H.}\ \bibnamefont
  {Boyer}},\ }\href {\doibase 10.1103/PhysRevD.11.790} {\bibfield  {journal}
  {\bibinfo  {journal} {Phys. Rev. D}\ }\textbf {\bibinfo {volume} {11}},\
  \bibinfo {pages} {790} (\bibinfo {year} {1975})}\BibitemShut {NoStop}%
\bibitem [{\citenamefont {Clerk}\ \emph {et~al.}(2010)\citenamefont {Clerk},
  \citenamefont {Devoret}, \citenamefont {Girvin}, \citenamefont {Marquardt},\
  and\ \citenamefont {Schoelkopf}}]{Clerk2010}%
  \BibitemOpen
  \bibfield  {author} {\bibinfo {author} {\bibfnamefont {A.~A.}\ \bibnamefont
  {Clerk}}, \bibinfo {author} {\bibfnamefont {M.~H.}\ \bibnamefont {Devoret}},
  \bibinfo {author} {\bibfnamefont {S.~M.}\ \bibnamefont {Girvin}}, \bibinfo
  {author} {\bibfnamefont {F.}~\bibnamefont {Marquardt}}, \ and\ \bibinfo
  {author} {\bibfnamefont {R.~J.}\ \bibnamefont {Schoelkopf}},\ }\href
  {\doibase 10.1103/RevModPhys.82.1155} {\bibfield  {journal} {\bibinfo
  {journal} {Rev. Mod. Phys.}\ }\textbf {\bibinfo {volume} {82}},\ \bibinfo
  {pages} {1155} (\bibinfo {year} {2010})}\BibitemShut {NoStop}%
\bibitem [{\citenamefont {Barrera}\ \emph {et~al.}(1985)\citenamefont
  {Barrera}, \citenamefont {Estevez},\ and\ \citenamefont
  {Giraldo}}]{Barrera1985}%
  \BibitemOpen
  \bibfield  {author} {\bibinfo {author} {\bibfnamefont {R.~G.}\ \bibnamefont
  {Barrera}}, \bibinfo {author} {\bibfnamefont {G.~A.}\ \bibnamefont
  {Estevez}}, \ and\ \bibinfo {author} {\bibfnamefont {J.}~\bibnamefont
  {Giraldo}},\ }\href {\doibase 10.1088/0143-0807/6/4/014} {\bibfield
  {journal} {\bibinfo  {journal} {European Journal of Physics}\ }\textbf
  {\bibinfo {volume} {6}},\ \bibinfo {pages} {287} (\bibinfo {year}
  {1985})}\BibitemShut {NoStop}%
\bibitem [{\citenamefont {Carrascal}\ \emph {et~al.}(1991)\citenamefont
  {Carrascal}, \citenamefont {Estevez}, \citenamefont {Lee},\ and\
  \citenamefont {Lorenzo}}]{Carrascal1991}%
  \BibitemOpen
  \bibfield  {author} {\bibinfo {author} {\bibfnamefont {B.}~\bibnamefont
  {Carrascal}}, \bibinfo {author} {\bibfnamefont {G.~A.}\ \bibnamefont
  {Estevez}}, \bibinfo {author} {\bibfnamefont {P.}~\bibnamefont {Lee}}, \ and\
  \bibinfo {author} {\bibfnamefont {V.}~\bibnamefont {Lorenzo}},\ }\href
  {\doibase 10.1088/0143-0807/12/4/007} {\bibfield  {journal} {\bibinfo
  {journal} {European Journal of Physics}\ }\textbf {\bibinfo {volume} {12}},\
  \bibinfo {pages} {184} (\bibinfo {year} {1991})}\BibitemShut {NoStop}%
\bibitem [{\citenamefont {Gittes}\ and\ \citenamefont
  {Schmidt}(1998)}]{Gittes1998}%
  \BibitemOpen
  \bibfield  {author} {\bibinfo {author} {\bibfnamefont {F.}~\bibnamefont
  {Gittes}}\ and\ \bibinfo {author} {\bibfnamefont {C.~F.}\ \bibnamefont
  {Schmidt}},\ }\href {\doibase 10.1364/OL.23.000007} {\bibfield  {journal}
  {\bibinfo  {journal} {Opt. Lett.}\ }\textbf {\bibinfo {volume} {23}},\
  \bibinfo {pages} {7} (\bibinfo {year} {1998})}\BibitemShut {NoStop}%
\bibitem [{\citenamefont {Glauber}(1963)}]{Glauber1963}%
  \BibitemOpen
  \bibfield  {author} {\bibinfo {author} {\bibfnamefont {R.~J.}\ \bibnamefont
  {Glauber}},\ }\href {\doibase 10.1103/PhysRev.130.2529} {\bibfield  {journal}
  {\bibinfo  {journal} {Phys. Rev.}\ }\textbf {\bibinfo {volume} {130}},\
  \bibinfo {pages} {2529} (\bibinfo {year} {1963})}\BibitemShut {NoStop}%
\bibitem [{\citenamefont {Carmichael}(1987)}]{Carmichael1987}%
  \BibitemOpen
  \bibfield  {author} {\bibinfo {author} {\bibfnamefont {H.~J.}\ \bibnamefont
  {Carmichael}},\ }\href {\doibase 10.1364/josab.4.001588} {\bibfield
  {journal} {\bibinfo  {journal} {Journal of the Optical Society of America B}\
  }\textbf {\bibinfo {volume} {4}},\ \bibinfo {pages} {1588} (\bibinfo {year}
  {1987})}\BibitemShut {NoStop}%
\end{thebibliography}%

\end{document}